\documentclass[twocolumn,floatfix,nofootinbib,showpacs,prd,aps,tightenlines,superscriptaddress]{revtex4-2}
\usepackage{amsmath}
\usepackage{graphicx}
\usepackage{psfrag}
\usepackage[dvipsnames]{xcolor}
\usepackage{dcolumn}
\usepackage{bm}
\usepackage[mathscr]{eucal}
\usepackage{mathrsfs}
\usepackage{xspace}
\usepackage{csquotes}
\usepackage{tabularx}
\newcolumntype{Z}{>{\centering\let\newline\\\arraybackslash\hspace{0pt}}X}
\usepackage{siunitx}
\usepackage[caption=false]{subfig}
\usepackage{tikz}
\usetikzlibrary{arrows,calc,decorations.markings,math,arrows.meta}
\usepackage{comment}
\usepackage{enumitem}
\usepackage{orcidlink}

\usepackage{hyperref}

\newcommand{\bea}{\begin{eqnarray}}
\newcommand{\eea}{\end{eqnarray}}
\newcommand{\beq}{\begin{equation}}
\newcommand{\eeq}{\end{equation}}

\newcommand{\ET}{{\sc Einstein Toolkit}\xspace}
\newcommand{\LAZEV}{{\sc LazEv}\xspace}
\newcommand{\Zforc}{{\sc Z4c}\xspace}
\newcommand{\Zforcow}{{\sc Z4cow}\xspace}
\newcommand{\ZforcowGPU}{{\sc Z4cowGPU}\xspace}
\newcommand{\SpacetimeX}{{\sc SpacetimeX}\xspace}
\newcommand{\TwoPunctureX}{{\sc TwoPunctureX}\xspace}

\newcommand{\Generato}{{\sc Generato}\xspace}
\newcommand{\Mathematica}{{\sc Mathematica}\xspace}
\newcommand{\Derivs}{{\sc Derivs}\xspace}
\newcommand{\CarpetX}{{\sc CarpetX}\xspace}
\newcommand{\Carpet}{{\sc Carpet}\xspace}
\newcommand{\Doutput}{{\it dense output}\xspace}

\definecolor{periwinkle}{RGB}{204, 204, 255}

\relax

\begin{document}

\def\fun#1#2{\lower3.6pt\vbox{\baselineskip0pt\lineskip.9pt
  \ialign{$\mathsurround=0pt#1\hfil##\hfil$\crcr#2\crcr\sim\crcr}}}
\def\lap{\mathrel{\mathpalette\fun <}}
\def\gap{\mathrel{\mathpalette\fun >}}
\def\kms{{\rm km\ s}^{-1}}
\def\vk{V_{\rm recoil}}
\def\araa{Annual Reviews of Astronomy \& Astrophysics}

\title{GPU-accelerated Subcycling Time Integration with the Einstein Toolkit}


\author{Liwei Ji\,\orcidlink{0000-0001-9008-0267}}
\email{ljsma@rit.edu}
\affiliation{Center for Computational Relativity and Gravitation, and School of Mathematical Sciences, Rochester Institute of Technology, 85 Lomb Memorial Drive, Rochester, New York 14623, USA}

\author{Roland Haas\,\orcidlink{0000-0003-1424-6178}}
\affiliation{Department of Physics \& Astronomy, University of British Columbia, Vancouver, Canada}
\affiliation{National Center for Supercomputing applications, University of Illinois, 1205 W Clark St, Urbana, IL, United States of America}
\affiliation{Department of Physics, University of Illinois, 1110 West Green St, Urbana, IL, United States of America}

\author{Yosef Zlochower\,\orcidlink{0000-0002-7541-6612}}
\affiliation{Center for Computational Relativity and Gravitation, and School of Mathematical Sciences, Rochester Institute of Technology, 85 Lomb Memorial Drive, Rochester, New York 14623, USA}

\author{Steven R Brandt\,\orcidlink{0000-0002-7979-2906}}
\affiliation{Center for Computation \& Technology, Louisiana State University, Baton Rouge, LA, United States of America}

\author{Erik Schnetter\,\orcidlink{0000-0002-4518-9017}}
\affiliation{Perimeter Institute for Theoretical Physics, Waterloo, Ontario, Canada}
\affiliation{Department of Physics and Astronomy, University of Waterloo, Waterloo, Ontario, Canada}
\affiliation{Center for Computation \& Technology, Louisiana State University, Baton Rouge, LA, United States of America}

\author{Allen Wen\,\orcidlink{0000-0001-9505-6557}}
\affiliation{Center for Computational Relativity and Gravitation, and School of Mathematical Sciences, Rochester Institute of Technology, 85 Lomb Memorial Drive, Rochester, New York 14623, USA}

\begin{abstract} 
Adaptive Mesh Refinement (AMR) with subcycling in time enables different grid levels to advance using their own time steps, ensuring finer grids employ smaller steps for accuracy while coarser grids take larger steps to improve computational efficiency. 
We present the development, validation, and performance analysis of a subcycling in time algorithm implemented within the \CarpetX driver in the \ET framework. This new approach significantly improves upon the previous subcycling implementation in the \Carpet driver by achieving higher-order convergence---fourth order in time instead of second order---and enhanced scaling performance. The key innovation lies in optimizing the exchange of ghost points at refinement boundaries, limiting it to the same number as those at inter-process boundaries using \Doutput from coarser levels, thereby reducing computational and communication overhead compared to the implementation in \Carpet, which required a larger number of buffer zones.

To validate the algorithm, we first demonstrate its fourth-order convergence using a scalar wave test. We then apply the algorithm to binary black hole (BBH) simulations, confirming its robustness and accuracy in a realistic astrophysical scenario. The results show excellent agreement with the well-established \LAZEV code. Scaling tests on CPU (Frontera) and GPU (Vista) clusters reveal significant performance gains, with the new implementation achieving improved speed and scalability compared to the \Carpet-based version.
\end{abstract}

\maketitle

\section{Introduction}\label{sec:introduction}

The Cactus Computational Toolkit~\cite{bona1998dimensionalnumericalrelativityhyperbolic, 809251, Goodale:2002a} has provided a free and open source framework for numerically studying fully general relativistic spacetimes since 1997. Since that time, many groups around the world have developed their own numerical relativity (NR) codes using the Cactus toolkit. With the breakthroughs in numerical relativity of 2005~\cite{Pretorius:2005gq, Campanelli:2005dd, Baker:2005vv}, these codes were finally able to evolve binary black hole (BBH) mergers from inspiral through ringdown. 
Since 2004, Cactus has incorporated mesh refinement capabilities through the Carpet driver~\cite{Schnetter:2003rb}, which employs moving boxes of nested mesh refinement. Over the years, Carpet has become the backbone of the Cactus ecosystem, enabling the generation of data for numerous scientific publications and leveraging vast amounts of computational resources on national infrastructures such as Teragrid, XSEDE, and ACCESS.

In recent years, the numerical simulation community has increasingly shifted toward GPU-based computing, which now serves as the primary source of floating-point operations (FLOPs) for high-performance simulations. In response, several efforts have been made to modernize the Cactus framework to leverage GPU hardware. The CarpetX driver~\cite{schnetter_2022_6131529,Shankar:2022ful,Kalinani:2024rbk}, the successor to Carpet, was released in 2023. Built on the AMReX library~\cite{Zhang2019,Zhang2021amrex}, CarpetX not only offers a more general and flexible mesh refinement scheme but also provides support for GPUs across all major vendors. However, several key features and tools essential for fully utilizing GPU capabilities remain to be reimplemented.

One such feature is subcycling in time, a technique that reduces computational costs by taking fewer time steps on coarser grids. Subcycling has long been a valuable tool for improving the efficiency of numerical simulations. However, its implementation has historically faced challenges implementing higher-order convergent schemes, particularly in updating intermediate stages of Runge-Kutta calculations at the boundaries of refined patches. Carpet addressed this issue using ``buffer zones"---extra grid zones that are progressively discarded as each Runge-Kutta stage is processed. While effective, this approach incurs significant memory and computational overhead, limiting its efficiency.

A breakthrough came with the work of~\cite{mongwane2015toward}, who introduced the concept of \Doutput~\cite{1988ZaMM...68..260S} into the evolution of the BSSN formulation of the Einstein field equations~\cite{PhysRevD.59.024007,PhysRevD.61.044012}, building on earlier developments by~\cite{mccorquodale2011high}. This approach has since been adopted by several groups, including~\cite{Palenzuela:2018sly}, who have implemented similar techniques in their codes.
The key innovation of this method lies in its replacement of the traditional reliance on buffer zones with a third-order polynomial in time, constructed from Runge-Kutta coefficients, to interpolate values at ghost zones during intermediate stages. By eliminating the need for buffer zones, this method offers a more computationally efficient and memory-effective alternative, marking a substantial improvement in the handling of refined grid boundaries in numerical simulations.

In implementing subcycling for CarpetX, we adopted the \Doutput approach~\cite{mongwane2015toward}. While this method introduces its own computational costs, it provides significant advantages over the traditional buffer zone technique, particularly in terms of memory efficiency and scalability. This innovation enhances the capabilities of the CarpetX driver, making it a more powerful tool for modern numerical relativity simulations.

In this paper, we document the implementation of subcycling in CarpetX, the integration of \Doutput, and the results of numerical tests and benchmarks that validate these new capabilities. Our work demonstrates the improved efficiency and higher order of convergence achieved by the CarpetX driver when using subcycling, paving the way for more advanced and scalable simulations in numerical relativity and beyond.

\section{Implementation Subcycling Within \CarpetX}\label{sec:implementation}

Here, we describe our implementation of an explicit fourth-order Runge-Kutta method (RK4) with subcycling, based on the works of~\cite{mccorquodale2011high,mongwane2015toward,Zhang2019}.
The key idea is to use \Doutput from the coarser level to interpolate and fill the ghost cells at the adaptive mesh refinement (AMR) boundary on the finer levels.

Consider an evolved state vector ${\bf y}$ governed by the evolution equation
\begin{equation}
\frac{d {\bf y}}{dt} = {\bf f}({\bf y}).
\end{equation}
This equation is subject to appropriate physical boundary conditions, which are essential for ensuring the system's behavior aligns with the underlying physics.
For example, in simulations involving wave propagation, these boundary conditions may include outgoing wave conditions at spatial infinity. 
Further details on the specific form of ${\bf f}({\bf y})$ and boundary conditions will be provided in subsequent sections.

The standard RK4 method updates ${\bf y}$ from time $t_n$ to $t_{n+1} = t_{n} + h$ through the following stages:
\begin{eqnarray}
  {\bf k}_1 &=& {\bf f}(t_n, {\bf Y}_1), \label{eq:rk1}\\
  {\bf k}_2 &=& {\bf f}(t_n+\frac{h}{2}, {\bf Y}_2),\\
  {\bf k}_3 &=& {\bf f}(t_n+\frac{h}{2}, {\bf Y}_3),\\
  {\bf k}_4 &=& {\bf f}(t_n+h, {\bf Y}_4),\\
  {\bf y}_{n+1} &=& {\bf y}_n + \frac{h}{6} ({\bf k}_1 + 2 {\bf k}_2 + 2 {\bf k}_3 + {\bf k}_4),\label{eq:rk5}
  \end{eqnarray}
where the intermediate stage vectors are computed as
\begin{eqnarray}
{\bf Y}_1 &=& {\bf y}_n,\\
{\bf Y}_2 &=& {\bf y}_n + \frac{h}{2} {\bf k}_1,\\
{\bf Y}_3 &=& {\bf y}_n + \frac{h}{2} {\bf k}_2,\\
{\bf Y}_4 &=& {\bf y}_n + h {\bf k}_3.
\end{eqnarray}

Because ${\bf f}$ implements finite differences, it cannot be evaluated directly on boundary points. At outer (physical) boundary points, the solution is constructed using physical boundary conditions. At AMR boundary points, the intermediate stage vectors ${\bf Y}_i$ are computed on the finer grid boundaries using \Doutput from the parent grid.

The \Doutput algorithm utilizes ${\bf k}_1$ through ${\bf k}_4$ to evaluate ${\bf y}$ at any point between $t_n$ and $t_{n+1}$ at the full order of accuracy of the underlying Runge-Kutta scheme.
To achieve this, we express ${\bf y}(t)$ in terms of a {\it progress parameter}, $\theta$, as
\begin{equation}
    {\bf y}(t_n + \theta h) = {\bf y}(t_n) + h \sum_{i=1}^4 b_i(\theta) {\bf k}_i + {\cal O}(h^4),
    \label{eq:denseoutput1}
\end{equation}
where the coefficients $b_i(\theta)$ are derived by performing a Taylor expansion in $h$ and matching terms:
\begin{align}
    b_1(\theta)&=\theta-\frac{3}{2}\theta^2+\frac{2}{3}\theta^3, \\
    b_2(\theta)&=b_3(\theta)=\theta^2-\frac{2}{3}\theta^3, \\
    b_4(\theta)&=-\frac{1}{2}\theta^2+\frac{2}{3}\theta^3.
\end{align}

The \Doutput formula can also be used to evaluate time derivatives at
intermediate steps:
\begin{align}
    \frac{d^{(m)}}{dt^{(m)}}{\bf y}(t_n+\theta h)
    &=\frac{1}{h^{(m-1)}}\sum_{i=1}^4{\bf k}_i
    \frac{d^{(m)}}{d\theta^{(m)}}b_i(\theta)+\mathcal{O}(h^{4-m}).
    \label{eq:denseoutput2}
\end{align}

%

The Taylor expansions of ${\bf Y}_i$ around ${\bf y}_n$ are given by:
\begin{align}
    {\bf Y}_1&={\bf y}_n, \label{eq:Y1} \\
    {\bf Y}_2&={\bf y}_n + \frac{h}{2}{\bf y}'_n, \label{eq:Y2} \\
    {\bf Y}_3&={\bf y}_n + \frac{h}{2}{\bf y}'_n + \frac{h^2}{4}{\bf y}''_n+\frac{h^3}{16}({\bf y}'''_n-{\bf f}_{\bf y} {\bf y}''_n) \label{eq:Y3} \\
    {\bf Y}_4&={\bf y}_n + h{\bf y}'_n + \frac{h^2}{2}{\bf y}''_n+\frac{h^3}{8}({\bf y}'''_n+{\bf f}_{\bf y} {\bf y}''_n), \label{eq:Y4}
\end{align}
where ${\bf f}_{\bf y}{\bf y}''_n=4({\bf k}_3-{\bf k}_2)/h^2$.
The key idea is to use \Doutput from the parent grid to evaluate ${\bf y}_n$, ${\bf y}'_n$, ${\bf y}''_n$ and ${\bf y}'''_n$
in the AMR boundary ghost zones
on the finer grid,
and then apply Eqs.~\eqref{eq:Y1}-\eqref{eq:Y4} to compute ${\bf Y}_i$ in the AMR ghost region.

Since the finer grid requires two steps to synchronize with the coarser grid,
we apply Eqs.~\eqref{eq:Y1}-\eqref{eq:Y4} twice to interpolate the boundary values $Y_i$ on the finer grid at $t_n$ ($\theta=0$) and $t_n+h/2$ ($\theta=1/2$).

We summarize the algorithm as follows. A superscript in parentheses denotes a given refinement level (i.e., ${\bf y}_n^{(\ell)}$ is the state vector on refinement level $\ell$ and $h^{(\ell)}$ is the grid spacing on that level).  Note that we define refinement levels such that $h^{(\ell)} / h^{(\ell+1)}=2.$

\begin{enumerate}

\item Perform a full update on the coarse level using the physical boundary condition to compute
${\bf Y}^{(0)}_1$, \dots, ${\bf Y}^{(0)}_4$
at the boundary.
Store
${\bf k}^{(0)}_1$, \ldots, ${\bf k}^{(0)}_4$.

\item On the next finer level $\ell$ prolongate in space
${\bf k}^{(\ell-1)}_1$, \ldots, ${\bf k}^{(\ell-1)}_4$
and
${\bf y}_n^{(\ell-1)}$
to mesh refinement boundary.

\item Perform a full update on level $\ell$. Use (\ref{eq:denseoutput1})-(\ref{eq:Y4}) with $\theta=0$ to compute
${\bf Y}^{(\ell)}_1$, \dots, ${\bf Y}^{(\ell)}_4$
at the mesh refinement boundary. 
Use (\ref{eq:denseoutput1}) with $\theta=1/2$ to compute
${\bf y}^{(\ell)}$
at mesh refinement boundary when implementing (\ref{eq:rk5}).
Store
${\bf k}^{(\ell)}_1$, \ldots, ${\bf k}^{(\ell)}_4$.

\item Recursively apply the algorithm to level $\ell+1$ starting at step 2.

\item Perform a full update on level $\ell$. Use (\ref{eq:denseoutput1})-(\ref{eq:Y4}) with $\theta=1/2$ to compute
${\bf Y}^{(\ell)}_1$, \dots, ${\bf Y}^{(\ell)}_4$
at the mesh refinement boundary. 
Use (\ref{eq:denseoutput1}) with $\theta=1$ to compute $y^{\ell}$ at mesh refinement boundary when implementing (\ref{eq:rk5}).
Store
${\bf k}^{(\ell)}_1$, \ldots, ${\bf k}^{(\ell)}_4$.

\item Recursively apply the algorithm to level $\ell+1$ starting at step 2.

\item Restrict to level $\ell-1$.

\end{enumerate}

As illustrated in Fig.~\ref{fig:bufferzone_vs_RMBghost}, \Carpet employs buffer zones to evolve fine grid points near AMR boundaries, requiring a prolongation width of four ghost zones for RK4 integration.
In contrast, \CarpetX utilizes the \Doutput method, which necessitates only a single ghost-zone-width prolongation.
\Carpet performs two prolongations of the state vector for every coarse time step, each involving $\sim$ 4 times as many points, while \CarpetX performs prolongation once per coarse time step but for 5 times as many variables (the ${\bf k}_1, \ldots, {\bf k}_4$ and the state vector). The net effect is \CarpetX prolongates  5/8 as many points.
In addition, because \Carpet treats three of the buffer zones as interior points, the RHS calculation is performed over more points (slowing the calculation).
The net effect is that \CarpetX is faster and scales better, as seen in Sec.~\ref{sec:tests:scaling}.

\begin{figure}

\resizebox{0.9\columnwidth}{!}{
\begin{tikzpicture}[scale=0.8]

    \def\delx{0.4}

    \def\xmin{0}
    \def\xmax{5}
    \def\ymin{0}
    \def\ymax{4}
    
    \foreach \y in {\ymin,...,\ymax} {
        \pgfmathtruncatemacro{\cond}{int(\y>\ymin && \y<\ymax)}
        \ifnum\cond=1
            \draw[gray,thin] (\xmin,\y) -- (\xmax+\delx,\y);
        \else
            \draw[gray,very thick] (\xmin,\y) -- (\xmax+\delx,\y);
        \fi
    }

    \def\csize{0.15}
    \foreach \x in {\xmin,...,\xmax} {
        \foreach \y in {\ymin,...,\ymax} {
            \pgfmathtruncatemacro{\cross}{int(\y>\x)}
            \ifnum\cross=1
                \draw
                    (\x-\csize,\y-\csize) -- (\x+\csize,\y+\csize)
                    (\x-\csize,\y+\csize) -- (\x+\csize,\y-\csize);
            \else
                \pgfmathtruncatemacro{\hollow}{int(\x<\xmax-1)}
                \ifnum\hollow=1
                    \draw (\x,\y) circle (4pt);
                \else
                    \fill (\x,\y) circle (4pt);
                \fi
            \fi
        }
    }
    
    \draw[gray,dashed, very thick]
        (\xmax-1.5,\ymin-0.2) -- (\xmax-1.5,\ymax+0.2);


    \def\xoffset{7}
    \pgfmathsetmacro{\xmin}{\xoffset}
    \pgfmathsetmacro{\xmax}{\xoffset + 5}
    \def\ymin{0}
    \def\ymax{4}
    
    \foreach \y in {\ymin,...,\ymax} {
        \pgfmathtruncatemacro{\cond}{int(\y>\ymin && \y<\ymax)}
        \ifnum\cond=1
            \draw[gray,thin] (\xmin,\y) -- (\xmax+\delx,\y);
        \else
            \draw[gray,very thick] (\xmin,\y) -- (\xmax+\delx,\y);
        \fi
    }
        
    \foreach \x in {\xmin,...,\xmax} {
        \foreach \y in {\ymin,...,\ymax} {
            \pgfmathtruncatemacro{\hollow}{int(\x<\xmin+1)}
            \ifnum\hollow=1
                \draw (\x,\y) circle (4pt);
            \else
                \fill (\x,\y) circle (4pt);
            \fi
        }
    }
    
    \draw[gray,dashed, very thick] (\xmin+0.5,\ymin-0.2) -- (\xmin+0.5,\ymax+0.2);

    \pgfmathsetmacro{\xdel}{0.8}
    \node at ({\xmax+\xdel}, \ymin) 
        [font=\Large, align=left, anchor=west] {previous time};
    \node at ({\xmax+\xdel}, \ymin+1) 
        [font=\Large, align=left, anchor=west] {first substep};
    \node at ({\xmax+\xdel}, \ymin+2) 
        [font=\Large, align=left, anchor=west] {second substep};
    \node at ({\xmax+\xdel}, \ymin+3) 
        [font=\Large, align=left, anchor=west] {third substep};
    \node at ({\xmax+\xdel}, \ymin+4) 
        [font=\Large, align=left, anchor=west] {final time};

\end{tikzpicture}
}

\caption{
Integration schemes at AMR boundaries: (left) \Carpet's buffer zone and (right) \CarpetX's ghost zone treatment.
The vertical dashed line separates interior points from AMR boundary zones.
Left (Carpet):
Empty circles ($\circ$) represent valid buffer zone points, while crosses ($\times$) represent points with invalid data during time integration. Each empty circle spans one ghost zone width, with each RK4 step requiring four ghost-zone-width prolongations.
Right (CarpetX):
Instead of excluding points, we prolongate the RK4 vectors ($\bf{y}_n, \bf{k}_1,...,\bf{k}_4$) across one ghost-zone width from the coarser level, then populate all ghost points at intermediate substeps and the final time using \Doutput method.
}
\label{fig:bufferzone_vs_RMBghost}

\end{figure}

\section{Physical system}

\subsection{Z4c Formalism}

We use the \Zforc~\cite{Bernuzzi:2009ex,Hilditch:2012fp} formulation of the Einstein evolution equations. The evolved quantities in this formulation are the conformal factor $W$, the conformal 3-metric $\tilde\gamma_{ij}$, the modified trace of the extrinsic curvature $\hat{K}=K-2\Theta$, the conformal trace-free part of the extrinsic curvature $\tilde{A}_{ij}$, the conformal contracted Christoffel symbols $\tilde\Gamma^i$, the constraint $\Theta$, the lapse $\alpha$, and the shift $\beta^i$. Their relation to the {\sc ADM}\xspace variables (the spatial metric $\gamma_{ij}$, the extrinsic curvature $K_{ij}$) is the following:
\begin{align}
    W
    &=\gamma^{-1/6},
    \\
    \tilde\gamma_{ij}
    &=W^2\gamma_{ij},
    \\
    \tilde{A}_{ij}
    &=W^2\left(K_{ij}-\frac{1}{3}\gamma_{ij}K\right),
    \\
    \tilde\Gamma^i
    &=-\partial_j\tilde\gamma^{ij}{}.
\end{align}
where $\gamma\equiv\det{\gamma_{ij}}$ and $K=\gamma^{ij}K_{ij}$.
Then equations of motion for the \Zforc formulation are given by:
\begin{align}
    \partial_tW
    &=\partial_\beta W
    +\frac{1}{3}W
    \left[
        \alpha\left(\hat{K}+2\Theta\right)-\partial_i\beta^i
    \right],
    \\
    \partial_t\tilde\gamma_{ij}
    &=\partial_\beta\tilde\gamma_{ij}
    -2\alpha\tilde{A}_{ij}
    +2\tilde\gamma_{k(i}\partial_{j)}\beta^k
    -\frac{2}{3}\tilde\gamma_{ij}\partial_k\beta^k,
    \\
    \partial_t\hat{K}
    &=\partial_\beta\hat{K}
    -D^iD_i\alpha
    +\alpha
    \left[
        \tilde{A}_{ij}\tilde{A}^{ij}+\frac{1}{3}(\hat{K}+2\Theta)^2
    \right]
    \nonumber \\
    &\quad
    +4\pi\alpha\left(S+\rho\right)
    +\alpha\kappa_1(1-\kappa_2)\Theta,
    \\
    \partial_t\tilde{A}_{ij}
    &=\partial_\beta\tilde{A}_{ij}
    +W^2
    \left[
        -D_iD_j\alpha
        +\alpha(R_{ij}-8\pi S_{ij})
    \right]^\text{TF}
    \nonumber\\
    &\quad
    +\alpha
    \left[
        (\hat{K}+2\Theta)\tilde{A}_{ij}
        -2\tilde{A}^k{}_i\tilde{A}_{kj}
    \right]
    \nonumber\\
    &\quad
    +2\tilde{A}_{k(i}\partial_{j)}\beta^k
    -\frac{2}{3}\tilde{A}_{ij}\partial_k\beta^k,
    \\
    \partial_t\tilde\Gamma^i
    &=\partial_\beta\tilde\Gamma^i
    -2\tilde{A}^{ij}\partial_j\alpha
    +2\alpha
    \bigg[
        \tilde\Gamma^i{}_{jk}\tilde{A}^{jk}
        -3\tilde{A}^{ij}\partial_j\ln W
        \nonumber\\
        &\quad
        -\frac{1}{3}\tilde\gamma^{ij}\partial_j(2\hat{K}+\Theta)
        -8\pi\tilde\gamma^{ij}S_j
    \bigg]
    +\tilde\gamma^{jk}\partial_j\partial_k\beta^i
    \nonumber\\
    &\quad
    +\frac{1}{3}\tilde\gamma^{ij}\partial_j\partial_k\beta^k
    -(\tilde\Gamma_d)^j\partial_j\beta^i
    +\frac{2}{3}(\tilde\Gamma_d)^i\partial_j\beta^j
    \nonumber\\
    &\quad
    -2\alpha\kappa_1
    \left[\tilde\Gamma^i-(\tilde\Gamma_d)^i\right],
    \\
    \partial_t\Theta
    &=\partial_\beta\Theta
    +\frac{1}{2}\alpha
    \left[
        R-\tilde{A}_{ij}\tilde{A}^{ij}
        +\frac{2}{3}(\hat{K}+2\Theta)^2
    \right]
    \nonumber\\
    &\quad
    -\alpha
    \left[
        8\pi\rho+\kappa_1(2+\kappa_2)\Theta
    \right]
\end{align}
where $D_i$ is the covariant derivative compatible with the ADM metric, and $\tilde{D}_i$ is the covariant derivative compatible with the conformal metric.
\begin{align}
    R_{ij}
    &=\tilde{R}^W{}_{ij}+\tilde{R}_{ij},
    \\
    \tilde{R}^W{}_{ij}
    &=\frac{1}{W}\tilde{D}_i\tilde{D}_jW
    +\frac{1}{W}\tilde\gamma_{ij}\tilde{D}^l\tilde{D}_lW
    \nonumber\\
    &\quad
    -2\tilde\gamma_{ij}\partial^l\ln W \partial_l\ln W,
    \\
    \tilde{R}_{ij}
    &=
    -\frac{1}{2}\tilde\gamma^{lm}\partial_l\partial_m\tilde\gamma_{ij}
    +\tilde\gamma_{k(i}\partial_{j)}\tilde\Gamma^k
    +(\tilde\Gamma_d)^k\tilde\Gamma_{(ij)k}
    \nonumber\\
    &\quad
    +\tilde\gamma^{lm}
    \left(
        2\tilde\Gamma^k{}_{l(i}\tilde\Gamma_{j)km}
        +\tilde\Gamma^k{}_{im}\tilde\Gamma_{klj}
    \right),
\end{align}
and $(\tilde\Gamma_d)^i=\tilde\gamma^{jk}\tilde\Gamma^i_{jk}$.

To complete the evolution system, we implement the puncture gauge conditions \cite{PhysRevLett.75.600, Alcubierre:2002iq, Campanelli:2005dd, PhysRevD.73.124011, PhysRevD.74.024016}: \begin{align}
    \partial_t\alpha
    &=\partial_\beta\alpha-2\alpha\hat{K},
    \label{eq:punclapse}
    \\
    \partial_t\beta^i
    &=\partial_\beta\beta^i
    +\frac{3}{4}\tilde\Gamma^i-\eta\beta^i.
    \label{eq:puncshift}
\end{align}
The function $\eta(r)$, which plays a critical role in the gauge condition, is defined as:
\begin{align}
    \eta(r)
    =(\eta_c-\eta_o)\exp (-(r/\eta_s)^4)+\eta_o
\end{align}
with parameters $\eta_c=2.0/M$, $\eta_s=40.0M$, and $\eta_o=0.25/M$.
This functional form ensures that $\eta$ is small in the outer regions of the computational domain.
As discussed in~\cite{Schnetter:2010cz}, the magnitude of $\eta$ imposes a constraint on the maximum allowable time step, such that $dt_\text{max}\propto1/\eta$. This constraint is independent of spatial resolution and becomes particularly relevant in the coarse outer zones. In these regions, the standard CFL condition would otherwise permit a significantly larger $dt_\text{max}$, but the presence of $\eta$ restricts the time step to ensure numerical stability.

\subsection{Initial Data}

The initial data for our simulations were generated using the \TwoPunctureX thorn \cite{Brandt:1997tf,Ansorg:2004ds} within the \SpacetimeX repository.
For the initial gauge conditions, we adopted the following form for the lapse function $\alpha$ and shift vector $\beta^i$:
\begin{align}
\alpha
&=\frac{1}{2}\left(1+\frac{1-\mathcal{M_+}/(2r_{+})-\mathcal{M_-}/(2r_{-})}{1+\mathcal{M_+}/(2r_{+})+\mathcal{M_-}/(2r_{-})}\right),
\\
\beta^i
&=0,
\end{align}
where $r_{\pm}$ is the coordinate distance to the respective black holes, and $\mathcal{M}_{\pm}$ represents the bare mass parameters of the two black holes. In this paper, we focus on the evolution of nonspinning binary black hole systems with mass ratio $q=1$. The initial data parameters were carefully chosen to align with the configurations described in~\cite{fernando2023massively}.

\subsection{Numerical Implementation}

Our implementation begins with the \Zforc thorn~\cite{Shankar:2022ful,Kalinani:2024rbk} in the \SpacetimeX  repository~\cite{spacetimex2024}. We replace the right-hand side (RHS) calculation with code generated by \Generato~\cite{generato2024}, a \Mathematica-based code generator. Additionally, we replace the derivatives with functions from the \Derivs thorn in the \CarpetX repository, extending it to support eighth-order finite differences and ninth-order Kreiss-Oliger (KO) dissipation.
For all spacetime simulations presented in this paper, we use eighth-order finite differences and fifth-order KO dissipation.

To fully leverage GPU clusters, we implement an additional thorn, \ZforcowGPU, and further optimize it. For simplicity, we remove the SIMD support~\cite{Shankar:2022ful,Kalinani:2024rbk} provided by the {\it nsimd} library~\cite{NSIMD}. We relocate all grid index calculations outside the GPU kernel, retaining only arithmetic operations within it. Instead of using the \Derivs thorn, we regenerate the finite difference stencils using \Generato, providing the option to either store the derivatives in temporary memory or compute them on-the-fly within the GPU kernels.

To formulate the system as a well-posed initial boundary value problem, outgoing boundary conditions~\cite{PhysRevD.67.084023} are applied to the dynamical variables through the {\sc NewRadX}~\cite{newradx2024} thorn in the \SpacetimeX repository.
These boundary conditions are crucial for preventing unphysical reflections and ensuring that the system evolves in a manner consistent with the underlying physics.

To stabilize numerical evolution, we incorporate
KO dissipation
~\cite{kreiss1973methods} into the system. This is implemented by modifying the time derivative of the evolved variables as follows:
\begin{align}
    \partial_tu
    \rightarrow
    \textsc{RHS}+Qu
\end{align}
where $Qu$ represents the dissipation term, given by:
\begin{align}
    Qu=(-1)^{r/2}\sigma/2^{r+2}
    \sum_ih_i^{r+1}D_{i+}^{r/2+1}D_{i-}^{r/2+1}u.
\end{align}
Here $r$ denotes the order of the finite differencing scheme used to evaluate \textsc{RHS}, $h_i$ is the grid spacing in the $i$-th direction, $D_{i+}$ and $D_{i-}$ are the forward and backward finite differencing operators, respectively. The parameter $\sigma$ controls the strength of the dissipation.

\section{Tests}\label{sec:tests}

In this section, we conduct a series of tests on the \Zforcow system with subcycling.

\subsection{Scalar Wave}

To validate the implementation of subcycling in \CarpetX, we first conducted tests using a linear wave equation governed by the following evolution equations:
\begin{align}
    \partial_tu&=\varPi,\\
    \partial_t\varPi&=\Delta u.
\end{align}
For the convergence analysis, the system was initialized with a Gaussian profile:
\begin{align}
    u(t,r)
    &=\frac{f(t-r)-f(t+r)}{r}, \\
    \varPi(t,r)
    &=-\frac{f(t-r)(t-r)-f(t+r)(t+r)}{\sigma^2r}
\end{align}
where $f(v)\equiv A \exp(-v^2/(2\sigma^2))$, with $A=1.0$, $\sigma=0.25/\sqrt{2}$.
The scalar wave equation was evolved at three different resolutions:
$\Delta x=h_1, h_1/1.2, h_1/1.44$, where $h_1=0.032$. The computational domain utilized a two-level refinement grid, with refinement boundaries at $r=4.0$, and $1.0$, respectively.

\begin{figure}
    \centering
    \includegraphics[width=\linewidth]{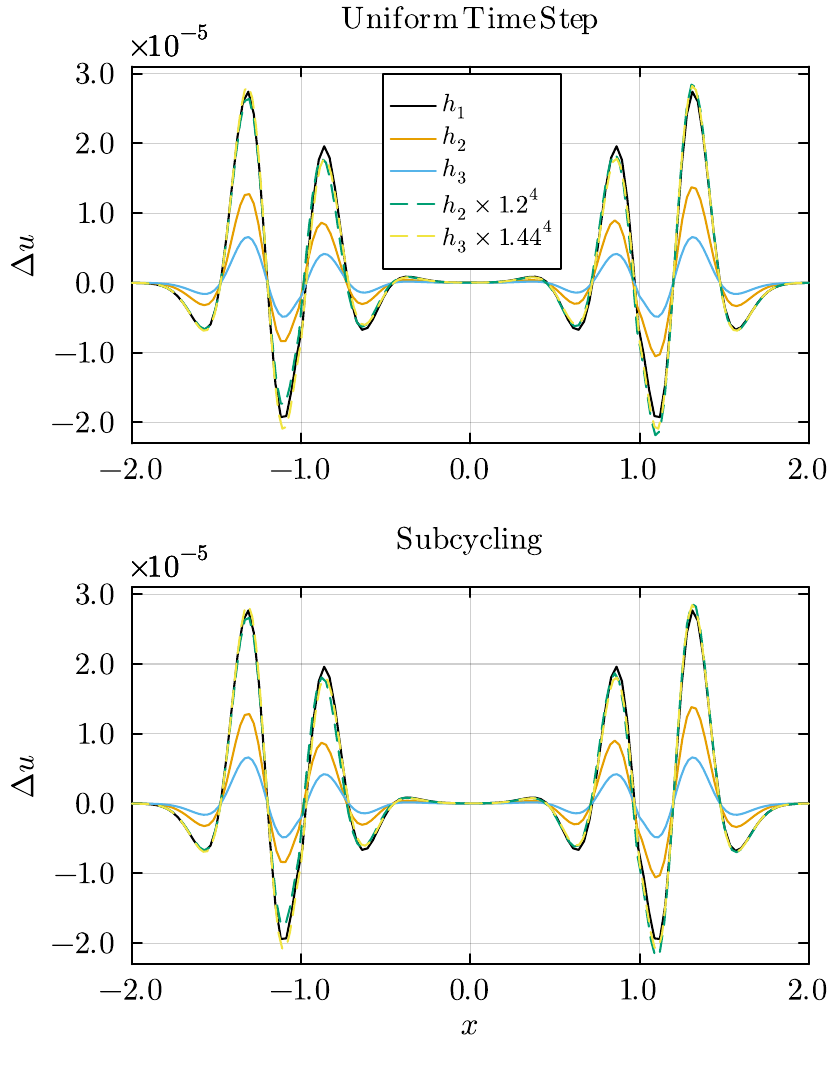}
    \caption{Errors in the solution of a Gaussian scaler wave. The plots show errors in
    low- ($h_1=0.032$),
    medium- ($h_2=h_1/1.2$),
    and high-resolution ($h_3=h_1/1.44$) numerical solutions for both uniform time step and subcycling cases. The dashed lines represent the scaled medium- and high-resolution errors assuming fourth-order convergence, and they align closely with the low-resolution solution.}
    \label{fig:ScalarWaveErrors}
\end{figure}

Figure \ref{fig:ScalarWaveErrors} displays the numerical errors at the second refinement level for three resolutions, comparing results with subcycling and uniform time stepping. The errors in the medium- and high-resolution cases were rescaled under the assumption of fourth-order convergence. The close alignment of these rescaled errors with the low-resolution curve confirms that the implementation achieves the expected fourth-order convergence, validating the accuracy and effectiveness of the subcycling approach.

\subsection{Binary Black Hole}\label{sec:tests:BBH}

To evaluate the correctness and accuracy of our implementation in a realistic astrophysical scenario, including subcycling in \CarpetX and the newly developed \Zforcow thorn within the \ET infrastructure~\cite{roland_haas_2024_14193969,Schnetter:2003rb,schnetter_2022_6131529,Brandt:1997tf,Ansorg:2004ds,Thornburg:2003sf,PhysRevD.65.044001}, we conducted a series of BBH simulations.
These tests included a comparison with the waveform generated by \LAZEV~\cite{Zlochower:2005bj, Campanelli:2005dd}.

The computational grid for these simulations consists of nine refinement levels, where $M$ represents the sum of the local ADM masses of individual black holes, computed in the asymptotically flat region at each puncture.
The simulation domain spans from $-400M$ to $400M$ in all three spatial dimensions, providing a sufficiently large volume to capture the dynamics of the BBH system while maintaining computational efficiency.

To facilitate direct comparisons with \LAZEV, we adopt an identical `box-in-box' mesh structure, with additional refinement levels defined by radii $r = 220$, $110$, $55$, $25$, $10$, $5$, $2$, and $1$.
This setup ensures consistency in the grid hierarchy and enables a precise assessment of the implementation's convergence behavior.

\subsubsection{Convergence tests for \Zforcow}

To evaluate the convergence properties of our new \Zforcow implementation, we evolve the $q=1$ binary system at three different resolutions: $\Delta x = h_1$, $h_1/1.2$, and $h_1/1.44$, where $h_1 = 3.3M$ represents the grid spacing on the coarsest level.

\begin{figure}
    \centering
    \includegraphics[width=\linewidth]{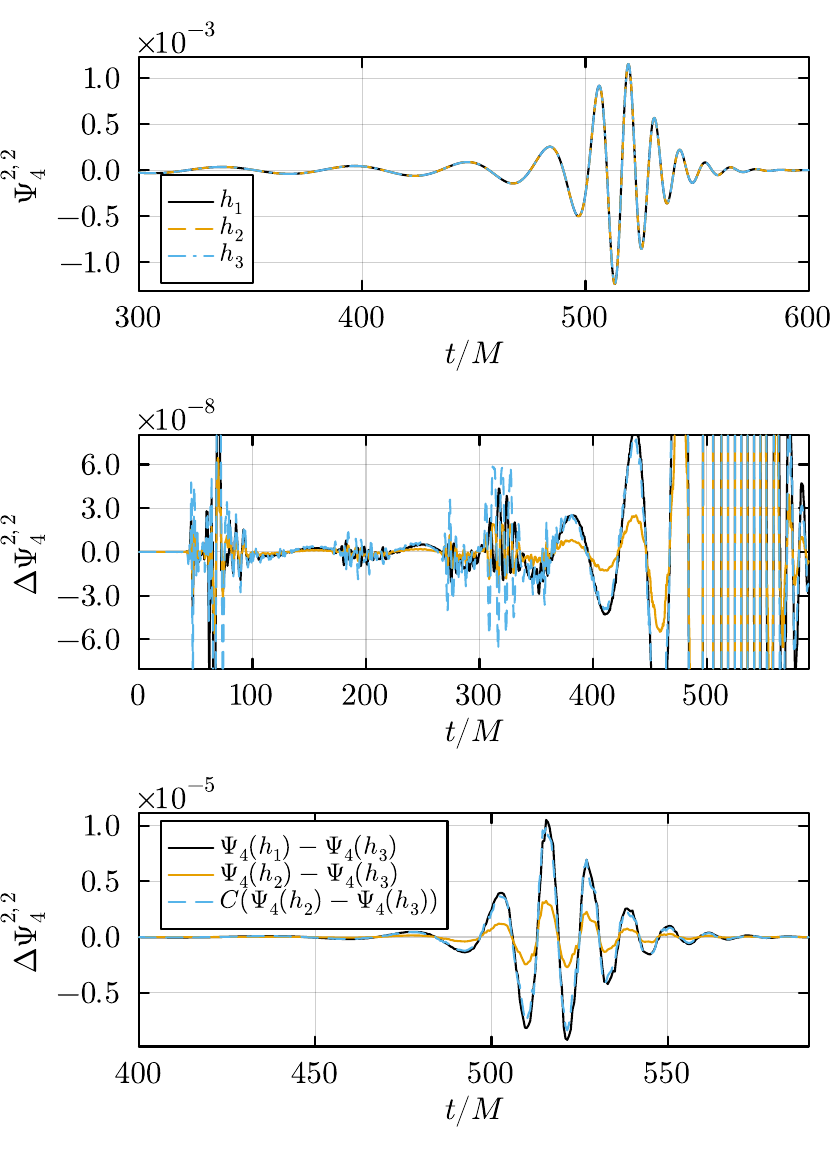}
    \caption{Convergence test of a $q=1$, nonspinning binary using \Zforcow. The top panel shows the
    low- ($h_1=3.3M$),
    medium- ($h_2=h_1/1.2$),
    and high-resolution ($h_3=h_1/1.44$) waveforms, which are visually indistinguishable, demonstrating excellent agreement across the resolutions.
    The middle and bottom panels display the waveform differences: the low- vs. high-resolution difference (solid black), the mid- vs. high-resolution difference (solid orange), and rescaled mid- vs. high-resolution difference (dashed blue), assuming fourth-order convergence. 
    }
    \label{fig:GW}
\end{figure}

In Fig.\ref{fig:GW}, the top panel illustrates the evolution of the real part of the
dominant
$l=2$, $m=2$ mode of the Weyl scalar $\Psi_4$, extracted at a radius of $r=60M$, which encodes outgoing gravitational radiation
through $\ddot{h}=\Psi_4$, where $h$ is the gravitational wave strain.
The evolution is shown over the time interval $t=300M$ to $t=600M$ for three different resolutions. 
Here, $\Psi_4$ is calculated using the {\sc WeylScal4} \cite{PhysRevD.65.044001} and {\sc Multipole} thorns in the \SpacetimeX repository.
At this scale, the results are visually indistinguishable, indicating excellent agreement across the resolutions.

The middle and bottom panels display the differences between the low- and high-resolution results, as well as the medium- and high-resolution results. Additionally, the difference between the medium- and high-resolution results is rescaled under the assumption of fourth-order convergence.
Numerical oscillations are observed around $t=70M$, $t=200M$, and $t=300M$, which are attributed to numerical errors that do not follow fourth-order convergence (middle panel). However, once the gravitational wave signal becomes stronger and dominates, starting around
$t=500M$
, the results exhibit clear fourth-order convergence (bottom panel). This behavior confirms the robustness and accuracy of the implementation in capturing the dominant physical features of the system.

\subsubsection{Comparison with \LAZEV}

\LAZEV, which evolves the BSSN formulation using the moving puncture approach~\cite{Campanelli:2005dd, Baker:2005vv}, serves as a benchmark for our analysis. Both \Zforcow and \LAZEV are integrated within the \ET infrastructure, ensuring a consistent framework for comparison.

To ensure a direct comparison with \LAZEV, we adopted the same initial lapse condition as \LAZEV in this section, specifically $\alpha(t=0)=\gamma^{-1/6}$, where $\gamma$ is the determinant of the 3-metric $\gamma_{ij}$. The \LAZEV simulations used for comparison here were originally published in~\cite{Fernando:2022php}.

\begin{figure}
    \centering
    \includegraphics[width=\linewidth]{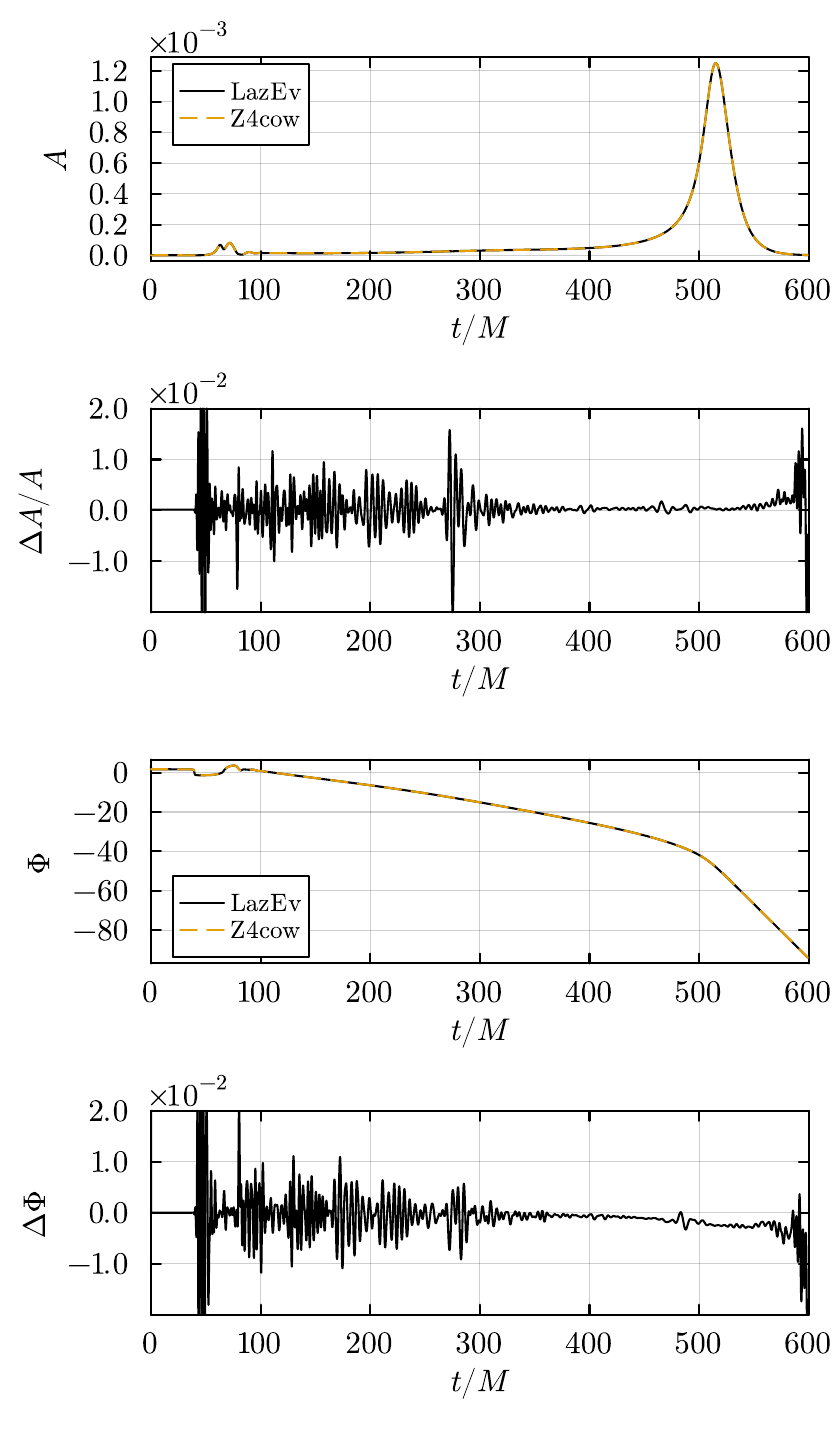}
    \caption{A comparison of the \Zforcow and \LAZEV  waveforms. The top panel shows the amplitudes of the waveforms from the two codes.
    The next panel down shows the relative differences in the amplitudes between the two codes. 
    The third panel show the waveform phases for the two codes. Finally, the bottom panel show the phase differences between the two codes.
    }
    \label{fig:GW-amp-phi}
\end{figure}

In Fig.~\ref{fig:GW-amp-phi}, we compare the waveforms from the \Zforcow and \LAZEV simulations. The top two panels show the amplitude of the $l=2,m=2$ mode of $\Psi_4$ for both \LAZEV and \Zforcow, along with their difference. The results demonstrate excellent agreement, with the maximum difference at merger being less than $0.1\%$. The bottom two panels present the phase of the $l=2,m=2$ mode of $\Psi_4$ for \LAZEV and \Zforcow, as well as their difference. The phase agreement is similarly excellent, with the phase difference at merger remaining below $0.001{\rm rad}$. These results highlight the high level of consistency between the two simulations, validating the accuracy and reliability of both the \Zforcow implementation and the subcycling approach.

\subsubsection{Irreducible mass}

\begin{figure}
    \centering
    \includegraphics[width=0.99\linewidth]{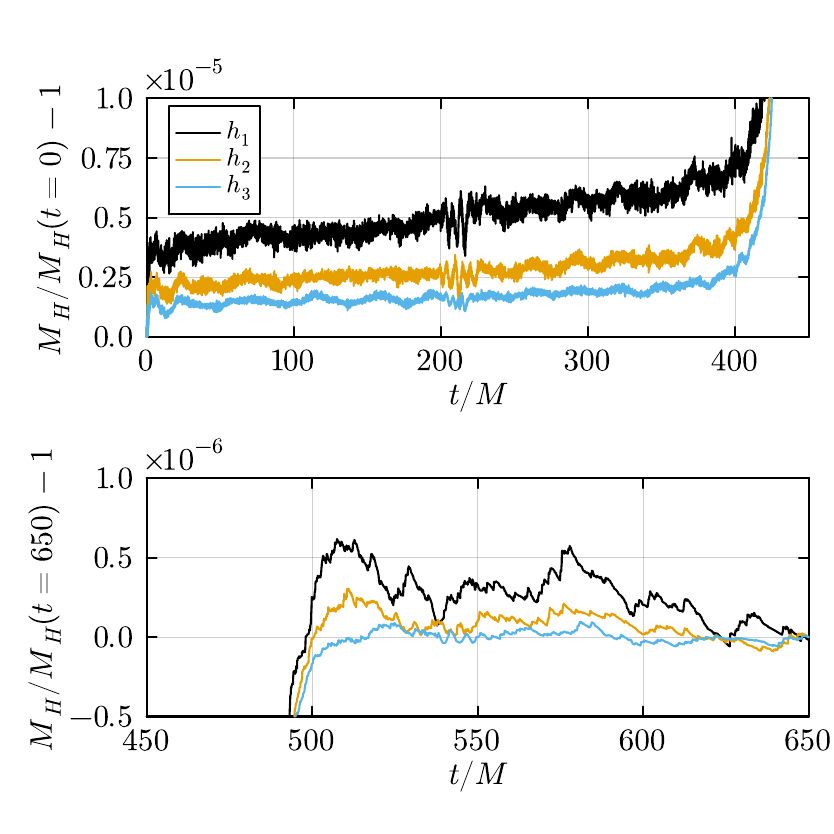}
    \caption{Irreducible mass of the inspiraling binary black holes (top) and the merged black hole (bottom), where $h_1=3.3M$, $h_2=h_1/1.2$, $h_3=h_1/1.44$.
    As the resolution increases, mass conservation improves.}
    \label{fig:Mass}
\end{figure}

In Fig.~\ref{fig:Mass} we present the evolution of the irreducible mass $M_\text{irr}$, calculated using the {\sc AHFinderDirect} thorn~\cite{Thornburg:2003sf} in the \SpacetimeX repository.
The top panel shows $M_\text{irr}$ for one of the inspiraling binary black holes (both exhibit identical behavior), while the bottom panel displays the irreducible mass of the post-merger black hole across three different resolutions. Ideally, $M_\text{irr}$ should remain constant during the inspiral and after merger. As expected, mass conservation improves progressively with increasing resolution.

We also observe a noise feature around $t=210M$, which we attribute to numerical errors at the refinement boundary between the coarsest and the second-coarsest levels. This noise feature diminishes with increasing resolution, indicating convergence and demonstrating the robustness of the numerical implementation at higher resolutions.

\subsubsection{Constraint violations}

\begin{figure}
    \centering
    \includegraphics[width=\linewidth]{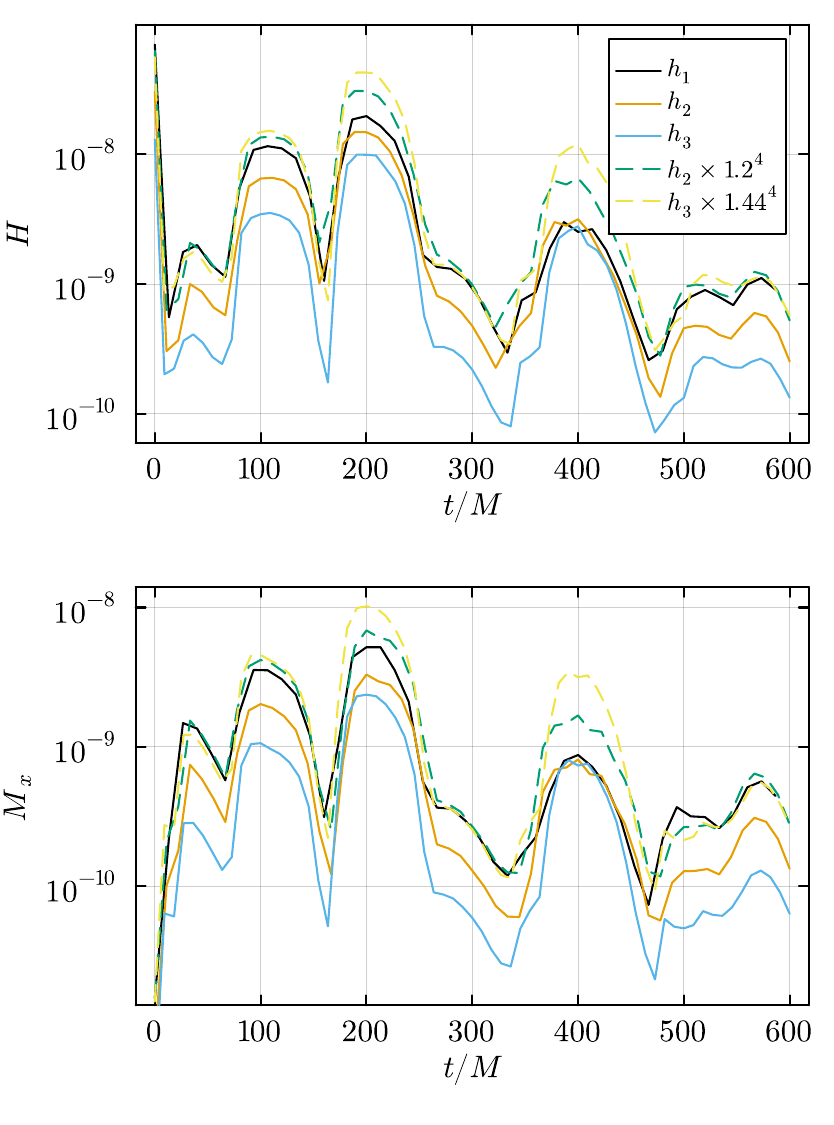}
    \caption{Hamiltonian and Momentum constraints for three different resolution cases, where $h_1=3.3M$, $h_2=h_1/1.2$, $h_3=h_1/1.44$.
    The dashed lines represent the scaled medium- and high-resolution constraints, assuming fourth-order convergence. Fourth-order convergence is observed, except around $t=200M$ and $t=400M$.}
    \label{fig:Constraints}
\end{figure}

Fig.~\ref{fig:Constraints} displays the $L_2$ norm of the Hamiltonian constraint and the $x$-component of the momentum constraint (the other two components exhibit similar behavior). The $L_2$ norm is computed over the region outside the two horizons (approximated using $\alpha = 0.3$) and within a sphere of radius $30M$.
The dashed lines are the scaled medium- and high-resolution constraint violations, assuming fourth-order convergence. The scaled constraints agree well with thee low-resolution case, except around $t=200M$ and $t=400M$.
We observe that the constraints at $t\sim 200M$ converge at lower order, while those at $t\sim 400M$ do not exhibit convergence.

\subsection{Scaling Test}\label{sec:tests:scaling}

\begin{figure}
    \centering
    \includegraphics[width=\linewidth]{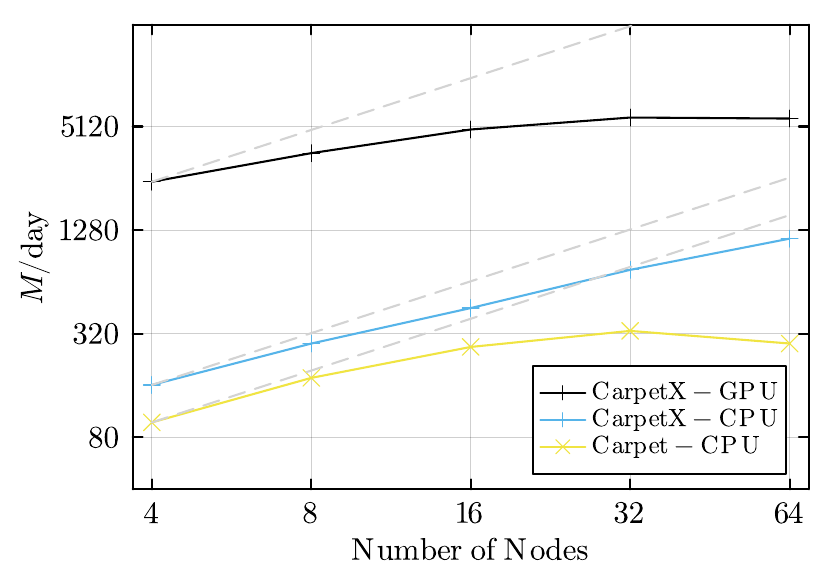}
    \caption{Strong scaling performance of \Zforcow thorn on CPU (Frontera) and GPU (Vista) clusters. The figure includes a \Carpet-compatible version of \Zforcow for comparison. The new subcycling algorithm, combined with the new \CarpetX driver, demonstrates improved computational speed and superior scaling compared to the \Carpet version. Additionally, significant performance gains are observed when running on GPU hardware.}
    \label{fig:comparecarpetandx}
\end{figure}

To demonstrate the efficiency of the new subcycling algorithms implemented in \CarpetX, we conducted scaling tests on both modern CPU (Frontera) and modern GPU (Vista) clusters, using the grid setup described in \ref{sec:tests:BBH}. For comparison with the older subcycling implementation in \Carpet, we generate a \Carpet-compatible version of \Zforcow using \Generato and performed scaling tests on Frontera. As shown in Fig.~\ref{fig:comparecarpetandx}, the \CarpetX version of \Zforcow exhibits both improved speed and better strong scaling compared to \Carpet version. Since the same RHS expressions are used in both implementations, the performance gains can be attributed to the enhanced subcycling algorithms in \CarpetX.

Additionally, with the new \CarpetX driver, we were able to run \Zforcow on a GPU machine (Vista). The scaling results from Vista, also displayed in Fig~\ref{fig:comparecarpetandx}, show significant performance improvements: the GPU executable runs approximately $15.2$ times faster for 4 nodes and $7.7$ times faster for 32 nodes compared to the CPU executable. These results highlight the substantial efficiency gains achieved by leveraging the new subcycling algorithms and GPU capabilities, underscoring the advancements enabled by \CarpetX in large-scale numerical simulations.

\begin{table*}
  \caption{Zone-cycles per second for the strong scaling tests, as reported in Fig.~\ref{fig:comparecarpetandx}.}
  \label{tab:scaling} 
  \begin{tabular}
  {rcc|rcc|rcc}
    \hline
    & Carpet CPU (Frontera) & &
    & CarpetX CPU (Frontera) & &
    & CarpetX GPU (Vista) & \\
    \hline
    CPUs & ZC/s & Efficiency &
    CPUs & ZC/s & Efficiency &
    GPUs & ZC/s & Efficiency \\
    \hline
    \hline
    $224$  & $5.88\times10^7$ & 1.00 &
    $224$  & $1.14\times10^8$ & 1.00 &
    $4$    & $1.82\times10^9$ & 1.00 \\
    $448$  & $1.07\times10^8$ & 0.91 &
    $448$  & $1.96\times10^8$ & 0.86 &
    $8$    & $2.60\times10^9$ & 0.71 \\
    $896$  & $1.61\times10^8$ & 0.68 &
    $896$  & $3.26\times10^8$ & 0.71 &
    $16$   & $3.48\times10^9$ & 0.48 \\
    $1792$ & $2.00\times10^8$ & 0.43 &
    $1792$ & $5.40\times10^8$ & 0.59 &
    $32$   & $4.03\times10^9$ & 0.28 \\
    $3584$ & $1.69\times10^8$ & 0.18 &
    $3584$ & $8.90\times10^8$ & 0.49 &
    $64$   & $3.97\times10^9$ & 0.14 \\
    \hline\hline 
  \end{tabular}
\end{table*}

To enable comparisons of zone cycle counts between simulations that do not use subcycling (e.g., {\sc Dendro-GR}~\cite{Fernando:2018mov,fernando2023massively}, {\sc AthenaK}~\cite{2024arXiv240916053S,Zhu:2024utz,Fields:2024pob}), we treat subcycling as a performance optimization. We define zone cycle counts {\it as if no subcycling had been used}, i.e., pretending that all grid cells take the same (smallest) time step size. While we acknowledge that this definition is not ideal, we argue that it serves as the most practical approach for the purposes of this section. The results are summarized in Table~\ref{tab:scaling} for the strong scaling tests and Table~\ref{tab:scaling-weak} for the weak scaling tests.

\begin{figure}
    \centering
    \includegraphics[width=\linewidth]{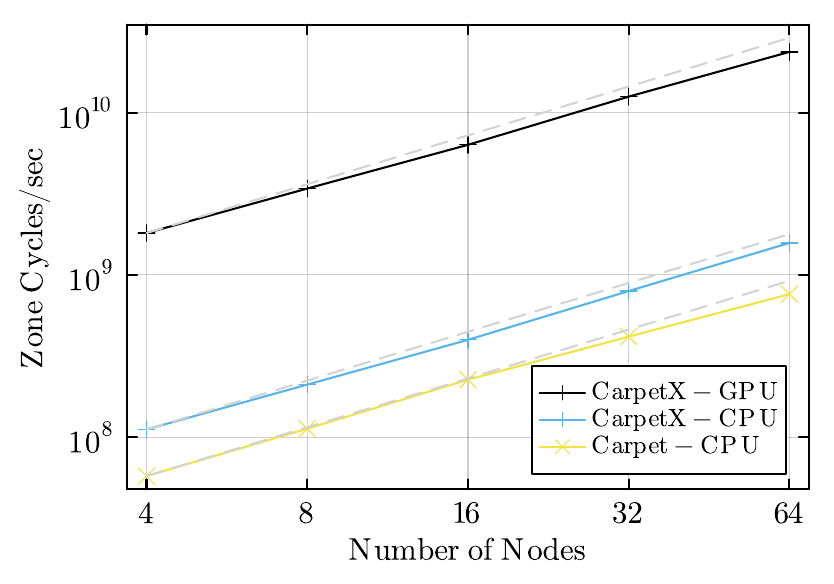}
    \caption{Weak scaling performance of \Zforcow thorn on CPU (Frontera) and GPU (Vista) clusters. The figure includes a \Carpet-compatible version of \Zforcow for comparison.
    The results demonstrate comparable weak scaling performance across all three test cases.
    }
    \label{fig:comparecarpetandx-weak}
\end{figure}

The weak scaling performance across the three test cases is comparable, with all cases demostraing good performance, as further illustrated in Fig.~\ref{fig:comparecarpetandx-weak}.
This performance highlights the robustness of the \CarpetX implementation and the new subcycling algorithm, showing their ability to maintain computational efficiency across different scaling regimes. Overall, these findings emphasize the significant advancements enabled by \CarpetX and the subcycling algorithm, which substantially enhance the performance of both CPU and GPU-based numerical simulations.
These developments pave the way for more efficient and scalable computational frameworks in numerical relativity, offering new opportunities for large-scale astrophysical simulations and broader applications in the field.

\begin{table*}
  \caption{Zone-cycles per second for the weak scaling tests, as reported in Fig.~\ref{fig:comparecarpetandx-weak}.}
  \label{tab:scaling-weak} 
  \begin{tabular}
  {rcc|rcc|rcc}
    \hline
    & Carpet CPU (Frontera) & &
    & CarpetX CPU (Frontera) & &
    & CarpetX GPU (Vista) & \\
    \hline
    CPUs & ZC/s/node & Efficiency &
    CPUs & ZC/s/node & Efficiency &
    GPUs & ZC/s/node & Efficiency \\
    \hline
    \hline
    $224$  & $1.44\times10^7$ & 1.00 &
    $224$  & $2.85\times10^7$ & 1.00 &
    $4$    & $4.55\times10^8$ & 1.00 \\
    $448$  & $1.41\times10^7$ & 0.98 &
    $448$  & $2.64\times10^7$ & 0.92 &
    $8$    & $4.26\times10^8$ & 0.94 \\
    $896$  & $1.41\times10^7$ & 0.98 &
    $896$  & $2.49\times10^7$ & 0.87 &
    $16$   & $3.96\times10^8$ & 0.87 \\
    $1792$ & $1.30\times10^7$ & 0.90 &
    $1792$ & $2.48\times10^7$ & 0.87 &
    $32$   & $3.94\times10^8$ & 0.87 \\
    $3584$ & $1.19\times10^7$ & 0.83 &
    $3584$ & $2.45\times10^7$ & 0.86 &
    $64$   & $3.70\times10^8$ & 0.81 \\
    \hline\hline 
  \end{tabular}
\end{table*}

\section{Discussion}\label{sec:discussion}

In this work, we have implemented a new subcycling algorithm within the \CarpetX driver in the \ET framework. Compared to the previous subcycling implementation in the \Carpet driver~\cite{Schnetter:2003rb}, our new approach offers higher-order convergence---fourth order instead of second order---and improved scaling performance. This improvement is achieved by limiting the exchange of ghost points at refinement boundaries to the same number as those at inter-process boundaries. In contrast, the old subcycling implementation required exchanging data in a buffer zone~\cite{Schnetter:2003rb}, which, in the case of RK4 integration, is four times larger.

We conducted rigorous testing to validate the new subcycling implementation in \CarpetX. First, we demonstrated fourth-order convergence using a scalar wave test, confirming the algorithm's accuracy and stability. Next, we applied the algorithm to a more complex and realistic scenario: BBH simulations. The results not only confirmed fourth-order convergence but also showed excellent agreement with the well-established \LAZEV code, highlighting the robustness and reliability of the new implementation.

Scaling tests on CPU (Frontera) and GPU (Vista) clusters further demonstrated the performance gains of the new implementation. Compared to the \Carpet-based version, the \CarpetX driver with subcycling achieves significantly better speed and scalability, making it a powerful tool for large-scale numerical relativity simulations. This improvement is particularly important for computationally demanding applications, such as BBH mergers, where efficiency and accuracy are critical.

While this work focuses on the implementation and testing of the subcycling algorithm for the RK4 method, extending it to RK2 and RK3 is straightforward. We plan to incorporate support for these methods in future work, further broadening the applicability and flexibility of the \CarpetX driver. These advancements represent a significant step forward in numerical relativity, enabling more efficient, accurate, and scalable simulations of complex astrophysical systems.

\begin{acknowledgments}

The authors would like to thank Manuela Campanelli, Michail Chabanov, Jay V. Kalinani, Carlos Lousto and Weiqun Zhang for valuable discussions.

The authors gratefully acknowledge the National Science Foundation (NSF) for financial support from grants OAC-2004044, OAC-2004157, OAC-2004879  PHY-2110338, OAC-2411068, PHY02409706, as well as the National Aeronautics and Space Administration (NASA) for financial support from TCAN Grant No. 80NSSC24K0100.
Research at Perimeter Institute is supported in part by the Government
of Canada through the Department of Innovation, Science and Economic
Development and by the Province of Ontario through the Ministry of
Colleges and Universities.
RH acknowledges additional support by NSF via grants OAC-2103680, OAC-2310548, and OAC-2005572.
ES and RH acknowledges the support of the Natural Sciences and Engineering
Research Council of Canada (NSERC). ES et RH remercie le Conseil de
recherches en sciences naturelles et en génie du Canada (CRSNG) de son
soutien.

This research used resources from the Texas Advanced Computing Center's (TACC) Frontera and Vista supercomputer allocations (award PHY20010). Additional resources were provided by the BlueSky, Green Prairies, and Lagoon clusters of the Rochester Institute of Technology (RIT) acquired with NSF grants PHY-2018420, PHY-0722703, PHY-1229173, and PHY-1726215.
We thank LSU HPC and CCT IT Services for making the \texttt{mike} compute resource available.


\end{acknowledgments}

\bibliographystyle{apsrev4-1}
\bibliography{ref}

\end{document}